# Speech Recognition of the letter 'zha'(ழ) in Tamil Language using HMM

A.Srinivasan*, K.Srinivasa Rao, K.Kannan and D.Narasimhan

Srinivasa Ramanujan Centre,SASTRA University,Kumbakonam-612001, Tamilnadu, India

**Abstract:**

Speech signals of the letter 'zha' (ழ) in Tamil language of 3 males and 3 females were coded using an improved version of Linear Predictive Coding (LPC). The sampling frequency was at 16 kHz and the bit rate was at 15450 bits per second, where the original bit rate was at 128000 bits per second with the help of wave surfer audio tool. The output LPC cepstrum is implemented in first order three state Hidden Markov Model(HMM) chain.

**Keywords**: Speech recognition, LPC, Tamil Language, Letter 'zha' (ழ), Wavesurfer, HMM.

## 1. Introduction

Tamil is one of the oldest and official languages in India. In Tamilnadu it is the prominent and primary language. It is one of the official languages of the union territories of Pondicherry and Andaman & Nicobar Islands. It is one of 23 nationally recognised languages in the Constitution of India. It has official status in Sri Lanka, Malaysia and Singapore. With more than 77 million speakers, Tamil is one of the widely spoken languages of the world. Tamil vowels are classified into short, long (five of each type) and two diphthongs. Consonants are classified into three categories with six in each category: hard, soft (a.k.a nasal), and medium. The classification is based on the place of articulation. In total there are 18 consonants. The vowels and consonants combine to form 216 compound characters. Placing dependent vowel markers on either one side or both sides of the consonant forms the compound characters. There is one more special letter *aytham* used in classical Tamil and rarely found in modern Tamil. In total there are 247 letters in Tamil alphabet. In these 247 letters Zha is the most significant, because of its usage and pronunciation. Many people will not pronounce the letter 'Zha' properly. There are two letters with same sound as Zha (la, lla), so it is necessary to recognize the letter Zha.

## 2.Vowels and Consonants

### 2.1 Vowels

There are 12 vowels in Tamil, called *uyireluttu* (*uyir* – life, *eluttu* – letter). These vowels are classified into short (*kuril*) and long (five of each type) and two diphthongs, /ai/ and /au/, and three "shortened" (*kurriyl*) vowels. The long vowels are about twice as long as the short vowels. The diphthongss are usually pronounced about 1.5 times as long as the short vowels.

|  | SHORT | | | LONG | | |
|---|---|---|---|---|---|---|
|  | **Front** | **Central** | **Back** | **Front** | **Central** | **Back** |
| **Close** | i |  | U | iː |  | iː |
|  | இ |  | உ | ஈ |  | ஊ |
| **Mid** | E |  | O | eː |  | oː |
|  | எ |  | ஒ | ஏ |  | ஓ |
| **Open** |  | A |  | (ai) | aː | (aw) |
|  |  | அ |  | ஐ | ஆ | ஔ |

**Table 1: Vowels in Tamil**

* Corresponding Author: A. Srinivasan, Department of ECE, Srinivasa Ramanujan Centre, SASTRA University, Kumbakonam-612001, Tamilnadu, India.  E-mail: asrinivasan78@yahoo.com

**2.2. Consonants**

Consonants are known as meyyeluttu (mey—body, eluttu—letters) in Tamil. It is classified into three categories with six in each category: vallinam (hard), mellinam (soft or Nasal) and itayinam (medium).Unlike most Indian languages, Tamil does not distinguish aspirated and unaspirated consonants. In addition, the voicing of plosives is governed by strict rules in centamil (Pure Tamil). Plosives are unvoiced if they occur word-initially or doubled. Elsewhere they are voiced, with a few becoming fricatives intervocalically. Nasals and approximants are always voiced.

As commonplace in languages of India, Tamil is characterised by its use of more than one type of coronal consonants. Retroflex consonants include the retroflex approximant /(H)/ (H) (example Tamil), which among the Dravidian languages is also found in Malayalam (example Kozhikode), disappeared from Kannada in pronunciation at around 1000 AD (the dedicated letter is still found in Unicode), and was never present in Telugu. Dental and alveolar consonants also contrast with each other, a typically Dravidian trait not found in the neighboring Indo-Aryan languages. In spoken Tamil, however, this contrast has been largely lost, and even in literary Tamil, e and d may be seen as allophonic.

A chart of the Tamil consonant phonemes in the International Phonetic Alphabet follows.Phonemes in brackets are voiced equivalents. Both voiceless and voiced forms are represented by the same character in Tamil, and voicing is determined by context. The sounds /f/ and /ʂ/ are peripheral to the phonology of Tamil, being found only in loanwords and frequently replaced by native sounds.

|  | Labial | Dental | Alveolar | Retroflex | Palatal | Velar |
|---|---|---|---|---|---|---|
| **Plosives** | p (b) | t̪ (d̪) |  | ʈ (ɖ) | tʃ (dʒ) | k (g) |
|  | ப | த |  | ட | ச | க |
| **Nasals** | M | n | N | ɳ | ɲ | Ŋ |
|  | ம | ந | ன | ண | ஞ | ங |
| **Tap** |  | ɾ |  |  |  |  |
|  |  | ர |  |  |  |  |
| **Trill** |  |  | R |  |  |  |
|  |  |  | ற |  |  |  |
| **Central Approximants** | ʋ |  |  | ɻ | j |  |
|  | வ |  |  | ழ | ய |  |
| **Lateral Approximants** |  | l |  | ɭ |  |  |
|  |  | ல |  | ள |  |  |

**Table 2: Consonants in Tamil**

**3. Analysis of 'Zha' using WaveSurfer**

WaveSurfer is a simple but powerful interface. The sound can be visualized and analyzed in several ways with the help of this tool. In addition, a spectrum window can be opened using Popup Spectrum Section for analyze Spectrum section plot (Magnitude Vs Frequency). Further the special control windows are available for Waveforms and Spectrograms, which allow the user to make quick modifications such as sound edit, noise elimination etc. The basic document we work with is sound files of 3 male and 3 female speakers with letter 'zha'. The standard speech analysis of the letter 'zha' such as Waveform and Spectrogram are analyzed and the samples are shown in following figures.

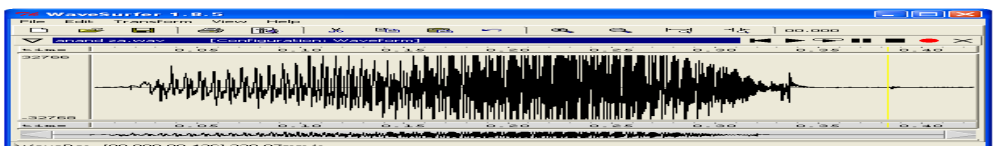

**Figure 1: Waveform of letter 'Zha' (ழ)**

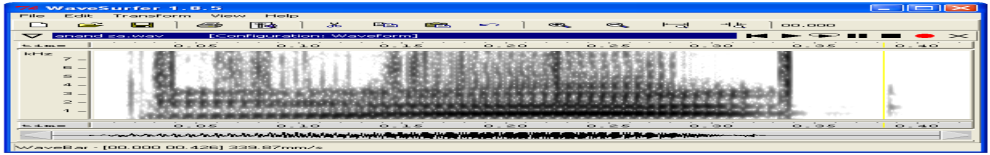

Figure 2: Spectrogram of letter 'Zha' (ழ)

### 4. Experimental Results

The variation of the Spectrum section plot in LPC is measures for the letter 'zha' is analyzed with the following parameters and the sample of magnitude Vs frequency plot is shown in figure 3

- Analysis type: LPC
- Analysis order: 20
- Speech signal bandwidth B = 8 kHz
- Sampling rate Fs = 16000 Hz (or samples/sec.)
- Channel: All
- Window type: Hamming
- Window length (frame): 512 points (20ms)
- Number of predictor coefficients of the LPC model = 18

### 4.1. Spectrum plot

Sample spectrum section plot of letter 'Zha' (ழ)

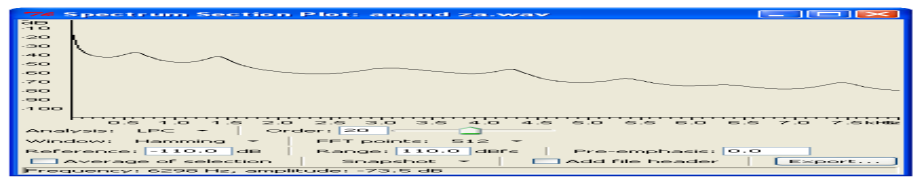

Figure 3: Sample spectrum section plot of letter 'Zha' (ழ)

### 4.2 Magnitude and frequency comparison of 3 male and 3 female speakers:

| Sl.No. | Frequency (Hz) | *F1 dB | *F2 dB | *F3 DB | **M1 dB | **M2 dB | **M3 dB |
|---|---|---|---|---|---|---|---|
| 1 | 15.625 | -20.47 | -20.98 | -19.67 | -21.02 | -21.86 | -20.71 |
| 2 | 140.625 | -32.68 | -32.73 | -32.01 | -32.94 | -33.13 | -32.75 |
| 3 | 390.625 | -36.63 | -36.74 | -36.13 | -36.94 | -37.14 | -37.01 |
| 4 | 640.625 | -41.36 | -34.99 | -40.97 | -41.48 | -41.92 | -40.99 |
| 5 | 1015.625 | -51.20 | -47.91 | -51.00 | -51.90 | -52.65 | -50.90 |
| 6 | 2046.875 | -51.68 | -53.12 | -51.02 | -52.02 | -52.98 | -53.05 |
| 7 | 3140.625 | -56.56 | -58.12 | -56.10 | -57.00 | -57.60 | -56.28 |
| 8 | 4203.125 | -61.15 | -68.51 | -61.12 | -61.66 | -61.98 | -63.22 |
| 9 | 5234.375 | -73.12 | -70.22 | -72.97 | -73.92 | -74.48 | -73.62 |
| 10 | 5703.125 | -72.42 | -68.77 | -72.27 | -72.95 | -73.25 | -72.99 |
| 11 | 6140.625 | -76.70 | -71.34 | -76.61 | -77.13 | -77.89 | -77.69 |
| 12 | 7171.875 | -73.45 | -72.08 | -73.39 | -73.99 | -74.19 | -73.16 |
| 13 | 7953.125 | -84.67 | -84.46 | -84.51 | -85.23 | -86.00 | -85.92 |
| 14 | 7984.375 | -84.74 | -84.52 | -84.57 | -85.56 | -86.02 | -85.83 |

*Female, ** Male

Table 3: Magnitude and frequency comparison of 3 male and 3 female speakers

## 5. Hidden Markov Model
### 5.1 HMM Formulation

A first-order N-state Markov chain for N = 3 is shown in Figure 4. The system can be described as being in one of the N distinct states 1,2, . . . , N at any discrete time instant t. We use the state variable qt as the state of the system at discrete time *t*. The Markov chain is then described by a state transition probability matrix A = [$a_{ij}$], where

$$a_{ij} = P(q_t = j | q_{t-1} = i), \quad 1 \leq i, j \leq N \quad (1)$$

with the following axiomatic constraints:

$$a_{ij} \geq 0 \text{ and} \quad (2)$$

$$\sum_{j=1}^{N} a_{ij} = 1 \text{ for all i.} \quad (3)$$

Note that in (1) we have assumed homogeneity of the Markov chain so that the transition probabilities do not depend on time. Assume that at t = 0 the state of the system $q_0$ is specified by an initial state probability $\pi_i$, = P ($q_0$ = i). Then, for any state sequence q = ($q_0$, $q_1$, $q_2$, . . . , $q_T$), the probability of q being generated by the Markov chain is

$$P(q | A, \pi) = \pi_{q_0} a_{q_0 q_1} a_{q_1 q_2} \ldots a_{q_{T-1} q_T} \quad (4)$$

Suppose now that the state sequence q cannot be readily observed. Instead, we envision each observation $O_t$, say a cepstral vector as mentioned previously, as being produced with the system in state $q_t$, $q_t \in \{1, 2, \ldots, N\}$. We assume that the production of $O_t$ in each possible state i (i = 1, 2, . . . N) is stochastic and is characterized by a set of observation probability measures B = $\{b_i(O_t)\}_{i=1}^{N}$,

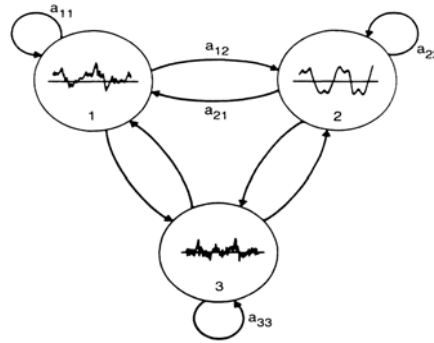

**Figure 4. A First-Order Three-State Markov Chain With Associated Processes.**

where

$$b_i(O_t) = P(O_t | q_t = i) \quad (5)$$

If the state sequence q that led to the observation sequence O = ($O_1$, $O_2$. . . , $O_T$) is known, the probability of 0 being generated by the system is assumed to be

$$P(O | \mathbf{q}, B) = b_{q_1}(O_1) b_{q_2}(O_2) \ldots b_{q_T}(O_T) \quad (6)$$

The joint probability of 0 and q being produced by the system is simply the product of (4) and (6), written as

$$P(o, q | \pi, A, B) = \pi_{q_0} \prod_{t=1}^{T} a_{q_{t-1} q_t} b_{q_t}(o_t) \quad (7)$$

It then follows that the stochastic process, represented by the observation sequence O, is characterized by

$$P(o | \pi, A, B) = \sum_{q} P(o, q | \pi, A, B)$$

$$= \sum_{q} \pi_{q_0} \prod_{t=1}^{T} a_{q_{t-1} q_t} b_{q_t}(o_t) \quad (8)$$

which describes the probability of observation being produced by the system without assuming the knowledge of the state sequence in which it was generated. The triple λ = (π, A, B) thus defines an HMM (8). In the following, we shall refer to λ as the model and the model parameter set interchangeably without ambiguity.

The particular formulation of (8) is quite similar to that of the incomplete data problem in statistics. In terms of the physical process of a speech signal, one interpretation that may be helpful for initial understanding of the problem is that a state represents an abstract speech code (such as a phoneme) embedded in a sequence of spectral observations, and because speech is normally produced in a continuous manner, it is often difficult and sometimes unnecessary to determine how and when a state transition (from one abstract speech code to another) is made. Therefore, in (8) we do not assume explicit, definitive observation of the state sequence **q,** although the Markovian structure of the state sequence is strictly implied. This is called a "Hidden Markov Model".

## 6. Statistical Method of HMM

In the development of the HMM methodology, the following problems are of particular interest. First, given the observation sequence O and a model λ, how do we efficiently evaluate the probability of O being produced by the source model λ-that is, P (0|λ) Second, given the observation O, how do we solve the inverse problem of estimating the parameters in λ. Although the probability measure of (8) does not depend explicitly on q, the knowledge of the most likely state sequence q that led to the observation (O) is desirable in many applications. The third problem then is how to deduce from observation the most likely state sequence q in a meaningful manner. According to convention, we call these three problems (i) the evaluation problem, (ii) the estimation problem and (iii) the decoding problem. In the following sections, we describe several conventional solutions to these three standard problems.

### 6.1 Evaluation Problem

The main concern in the evaluation problem is computational efficiency. Without complexity constraints, one can simply evaluate P (O|λ) directly from the definition of (8). Since the summation in (8) involves $N^{T+1}$ possible q sequences, the total computational requirements are on the order of 2T. $N^{T+1}$ operations. The need to compute (8) without the exponential growth of computation, as a function of the sequence length T, is the first challenge for implementation of the HMM technique. Fortunately, using the well-known forward-backward procedure, this exorbitant computational requirement of the direct summation can be easily alleviated. A forward induction procedure allows evaluation of the probability P (O|λ) to be carried out with only a computational requirement linear in the sequence length T and quadratic in the number of states N. To see how this is done, let us define the forward variable $\alpha_t(i)$ as $\alpha_t(i) = P(O_1, O_2, \ldots, O_t, q_t = i|\lambda)$ that is, the probability of the partial observation sequence up to time *t* and state $q_t = i$ at time *t*. With reference to Figure 5, which shows a trellis structure implementation of the computation of $\alpha_t(i)$, we see that the forward variable can be calculated inductively by

$$\alpha_t(j) = \left[ \sum_{i=1}^{N} \alpha_{t-1}(\mathbf{i}) \mathbf{a}_{ij} \right] b_j(o_t).$$

The desired result is simply $P(o|\lambda) = \sum_{i=1}^{N} \alpha_T(i)$. This tremendous reduction in computation makes the HMM method attractive and viable for speech recognizer designs because the evaluation problem can be viewed as one of scoring how well an unknown observation sequence (corresponding to the speech to be recognized) matches a given model (or sequence of models) source, thus providing an efficient mechanism for classification.

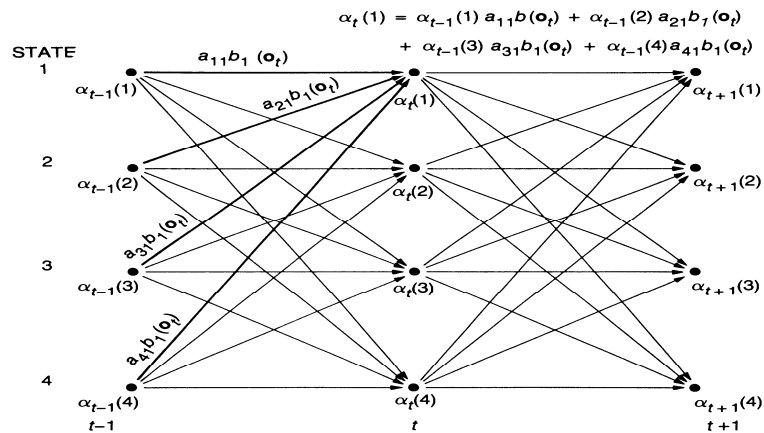

**Figure 5. A Trellis Structure for the Calculation of the Forward Partial Probabilities $\alpha_t(i)$.**

### 6.2 Estimation Problem

Given an observation sequence (or a set of sequences) O, the estimation problem involves finding the "right" model parameter values that specify a model most likely to produce the given sequence. In speech recognition, this is often called "training," and the given sequence, on the basis of which we obtain the model parameters, is called the training sequence, even though the formulation here is statistical. In solving the estimation problem, the method of maximum likelihood (ML); that is, we choose λ such that P (O|λ), as defined by (8), is maximized for the given training sequence O.

### 6.3 Decoding Problem

As noted previously, we often are interested in uncovering the most likely state sequence that led to the observation sequence O. Although the probability measure of an HMM, by definition, does not explicitly involve the state sequence, it is important in many applications to have the knowledge of the most likely state sequence for several reasons. As an example, if we use the states of a word model to represent the distinct sounds in the word, it may be desirable to know the correspondence between the speech segments and the sounds of the word, because the duration of the individual speech segments provides useful information for speech recognition. As with the second problem, there are several ways to define the decoding objective. The most trivial choice is, following the Bayesian framework, to maximize the (instantaneous) a posteriori probability.

### 6.4 Speech Recognition

The typical use of HMM's in speech recognition is not very different from the traditional pattern matching paradigm. Successful application of HMM methodsusually involves the following steps

1. Define a set of L sound classes for modeling, such as phonemes or words; call the sound classes V= $\{v_l, v_2, \ldots, v_L\}$.
2. For each class, collect a sizable set (the training set) of labeled utterances that are known to be in the class.
3. Based on each training set, solve the estimation problem to obtain a "best" model $\lambda_i$ for each class $v_i$ (i = 1, 2, . . . , L).
4. During recognition, evaluate P (O$|\lambda$) (i = 1, 2, . . ., L) for the unknown utterance observation and identify the speech that produced O as class $v_j$ if

$$P(o|\lambda_j) = \max_{1 \leq i \leq L} P(o|\lambda_i).$$

### 7. Conclusion

It is observed from voice excited LPC with Wavesurfer tool, there is a variation in magnitude of the letter 'Zha' among different people and result is implemented in first order Three-State Markov Chain With Associated Processes. To strengthen the results, more samples could collected from TamilNadu, Srilanka and Malasiya infuture and form a statistical methodology. This statistical methodology is lie in mathematical framework and implementational structure.